\documentclass[final,5p,times,twocolumn]{elsarticle}

\usepackage{amsfonts}
\usepackage{amsmath}
\usepackage{amssymb}
\usepackage{bm}
\usepackage{xcolor}
\usepackage{booktabs}
\usepackage[normalem]{ulem}
\usepackage[utf8]{inputenc}
\usepackage[english]{babel}
\biboptions{comma,sort&compress}

\usepackage{amsmath}
\usepackage{graphicx}
\usepackage[colorinlistoftodos]{todonotes}
\usepackage[colorlinks=true, allcolors=blue]{hyperref}
\usepackage[normalem]{ulem} 

\def\nuc#1#2{\relax\ifmmode{}^{#1}{\protect\text{#2}}\else${}^{#1}$#2\fi}

\begin{document}

\begin{frontmatter}



\title{Constraining the $^{12}$C+$^{12}$C astrophysical S-factors with the $^{12}$C+$^{13}$C measurements at very low energies}

\author[label1,label2]{N.T. Zhang}
\author[label1,label2]{X.Y. Wang}
\author[label1,label2]{H. Chen}
\author[label1,label2]{Z.J. Chen}
\author[label1]{W.P. Lin}
\author[label1,label2]{W.Y. Xin}
\author[label1]{S.W. Xu}
\author[label3,label4]{D. Tudor}
\cortext[mail]{Corresponding author}
\ead{dana.tudor@nipne.ro}
\author[label3,label4]{A.I. Chilug}
\author[label3,label4]{I.C. Stefanescu}
\author[label3]{M. Straticiuc}
\author[label3]{I. Burducea}
\author[label3]{D.G. Ghita}
\author[label3]{R. Margineanu}
\author[label3]{C. Gomoiu}
\author[label3]{A. Pantelica}
\author[label3]{D. Chesneanu}
\author[label3]{L. Trache}
\ead{livius.trache@nipne.ro}
\author[label1,label2,label5]{X.D. Tang}
\ead{xtang@impcas.ac.cn}
\author[label6]{B. Bucher}
\author[label7]{L.R. Gasques}
\author[label8,label9]{K. Hagino}
\author[label1,label10]{S. Kubono}
\author[label11]{Y.J. Li}
\author[label11]{C.J. Lin}
\author[label12]{A.S. Umar}
\author[label13]{Y. Xu}
\address[label1]{Institute of Modern Physics, Chinese Academy of Sciences, Lanzhou 730000, China}
\address[label2]{School of Nuclear Science and Technology, University of Chinese Academy of Sciences, Beijing 100049, China}
\address[label3]{Horia Hulubei National Institute of Physics and Nuclear Engineering, IFIN-HH, M$\breve{a}$gurele 077125, Romania}
\address[label4]{Doctoral School of Physics, University of Bucharest, Bucuresti-M$\breve{a}$gurele 077125, Romania}
\address[label5]{Joint department for nuclear physics, Lanzhou University and Institute of Modern Physics, Chinese Academy of Sciences, Lanzhou 730000, China}
\address[label6]{Idaho National Laboratory, Idaho Falls, Idaho 83415, USA}
\address[label7]{Departamento de F\'{i}sica Nuclear, Instituto de F\'{i}sica da Universidade de S\~{a}o Paulo, 05508-090, S\~{a}o Paulo, SP, Brazil}
\address[label8]{Department of Physics, Tohoku University, Sendai 980-8578, Japan}
\address[label9]{Research Center for Electron Photon Science, Tohoku University, 1-2-1 Mikamine, Sendai 982-0826, Japan}
\address[label10]{RIKEN Nishina Center, RIKEN, 2-1 Hirosawa, Saitama 351-0198, Japan}
\address[label11]{China Institute of Atomic Energy, Beijing 102413, China}
\address[label12]{Department of Physics and Astronomy, Vanderbilt University, Nashville, Tennessee 37235, USA}
\address[label13]{Extreme Light Infrastructure-Nuclear Physics, M$\breve{a}$gurele 077125, Romania}


\begin{abstract}
We use an underground counting lab with an extremely low background to perform an activity measurement for the $^{12}$C+$^{13}$C system with energies down to $E\rm_{c.m.}$=2.323 MeV, at which the $^{12}$C($^{13}$C,$p$)$^{24}$Na cross section is found to be 0.22(7) nb. The $^{12}$C+$^{13}$C fusion cross section is derived with a statistical model calibrated using experimental data. Our new result of the $^{12}$C+$^{13}$C fusion cross section is the first decisive evidence in the carbon isotope systems which rules out the existence of the astrophysical S-factor maximum predicted by the phenomenological hindrance model, while confirming the rising trend of the S-factor towards lower energies predicted by other models, such as CC-M3Y+Rep, DC-TDHF, KNS, SPP and ESW. After normalizing the model predictions with our data, a more reliable upper limit is established for the $^{12}$C+$^{12}$C fusion cross sections at stellar energies.
\end{abstract}


\end{frontmatter}


\section{Introduction}

The $^{12}$C+$^{12}$C fusion reaction is one of the most important reactions in nuclear astrophysics. The crucial energy range extends from a few 10's keV to $E\rm_{c.m.}$=3 MeV~\cite{Gasques2007}. All the measurements, going back 60 years, have been limited to energies above $E\rm_{c.m.}$=2.1 MeV. The situation is further complicated by the existence of resonance states with $\sim$50 keV widths and a few hundred keV spacings, continuing down to the lowest energies for which measurements have been made~\cite{Barnes+Tren+Wu}. Extrapolation is inevitably necessary to obtain the reaction rate for astrophysical applications.

\begin{figure}[h]
\includegraphics[scale=0.44]{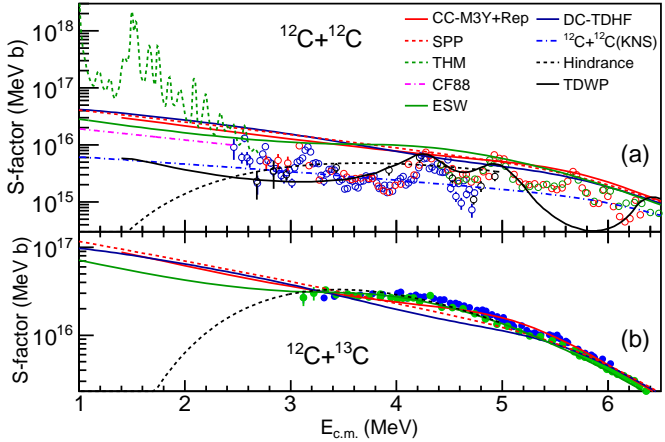}
\caption{\label{fig1}(Color online) S-factors of $^{12}$C+$^{12}$C (upper panel) and $^{12}$C+$^{13}$C (lower panel) at sub-barrier energies. The $^{12}$C+$^{12}$C data from Ref.~\cite{Becker1981}, \cite{Aguilera2006-p64601}, \cite{Spillane2007-p122501}, and \cite{Jiang2018} with errors better than 40\% are shown as red, green, blue and black open circles, respectively. The $^{12}$C+$^{13}$C data from Ref.~\cite{Dayras1976} and \cite{Dasmahapatra1982-p257} are shown respectively as green and blue filled circles. Model calculations, DC-TDHF (dark blue), CC-M3Y+Rep (red), TDWP (black) and CF88 (magenta), are shown as solid lines; SPP using global density distribution (red) and the proposed hindrance signatures (black) are shown as the dashed lines. The KNS prediction for the $^{12}$C+$^{12}$C non-resonant component and the THM measurement are shown as the light blue dashed-dotted line and dark green dashed line, respectively.}
\end{figure}

Modeling $^{12}$C+$^{12}$C heavy ion fusion cross sections at deep sub-barrier energies has been a long standing challenge for nuclear reaction theory~\cite{Barnes+Tren+Wu}. The Gamow energy window for most astrophysical applications typically spans 500 keV or more, and with no reliable method for predicting the resonances at lower energies, the standard reaction rate (CF88) was established by using constant modified S-factors ($S^*(E)=S(E)exp(0.46E)$) based on the square well penetration factor~\cite{Patterson1969,CF88}. This rising trend of S-factor towards lower energies is confirmed by various phenomenological and microscopic models, such as density-constrained time dependent Hartree-Fock method (DC-TDHF)~\cite{Umar2012}, wave-packet dynamics (TDWP)~\cite{aw2018}, barrier penetration model based on the global S\~ao Paulo potentials (SPP)~\cite{Gasques2007} or the Krappe-Nix-Sierk potential (KNS)~\cite{Aguilera2006-p64601}, and coupled channel calculations such as CC-M3Y+Rep~\cite{des2013,Esbensen2011} (Fig.~\ref{fig1}(a)). However, the hindrance model, a global phenomenological model based on the systematics observed in systems with 64$\gtrsim$A$\gtrsim$30, predicts that the $^{12}$C+$^{12}$C S-factor reaches its maximum at $E\rm_{c.m.}$=3.68$\pm$0.38 MeV~\cite{Back2014,Jiang2007,Jiang2018}. At lower energies, this model predicts a rapid drop in the S-factor leading to a reduced reaction rate that is many orders of magnitude smaller than the standard rate used for astrophysical modeling. Although the experimental data are available down to 2.1 MeV, no decisive conclusion can be made due to the poor fitting of the resonance-like structure in the $^{12}$C+$^{12}$C cross section with the smooth extrapolating models~\cite{Jiang2018}. Recently, the cross sections were determined indirectly for $E\rm_{c.m.}$= 0.8 MeV to 2.7 MeV using the Trojan Horse Method~\cite{thm2018}, recommending a new S-factor with a rising slope faster than any models presented in Fig.~\ref{fig1}(a), resulting in a new rate which is 1 or 2 orders of magnitude higher than the standard one. The large uncertainty of the $^{12}$C+$^{12}$C rate drastically impacts a number of models such as late-time massive star evolution, the ignition dynamics of type Ia supernovae and x-ray superbursts~\cite{Gasques2007,Kanji2018}.

\begin{figure}[h]
\includegraphics[scale=0.44]{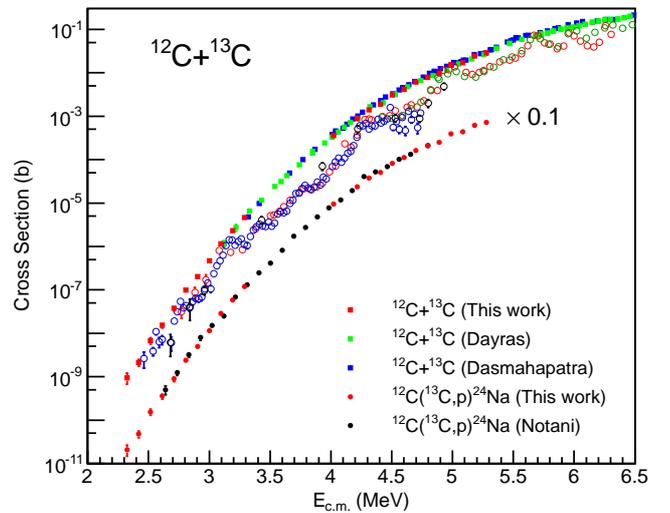}
\caption{\label{na24_xsec}(Color online) Cross sections of $^{12}$C($^{13}$C,$p$)$^{24}$Na are obtained in Ref.~\cite{Notani2012} (black) and this work (red). The latter represent the first measurements to reach the energy region E$\rm_{c.m.}$$<$2.6 MeV. A 10\% experimental systematical uncertainty is included in both data. The $^{12}$C+$^{13}$C total fusion cross sections from Ref.~\cite{Dayras1976}, \cite{Dasmahapatra1982-p257} are also shown as green and blue squares, respectively. The $^{12}$C+$^{12}$C total fusion cross sections are also shown as open circles with symbols explained in the Fig.~\ref{fig1} caption.}

\end{figure}

While the complicated resonance-like structure in $^{12}$C+$^{12}$C and the lack of reliable measurements at lower energies prevent us from drawing a clear conclusion~\cite{Jiang2018}, the isotope fusion reaction $^{12}$C+$^{13}$C offers an ideal opportunity to constrain the $^{12}$C+$^{12}$C S-factor. It has been observed at energies below and above the Coulomb barrier that $^{12}$C+$^{13}$C and $^{13}$C+$^{13}$C cross sections upper bound the $^{12}$C+$^{12}$C cross sections, and match the maxima of the resonance-like structure seen in $^{12}$C+$^{12}$C in a wide range from 10$^{-8}$ b to 1 b~\cite{Notani2012} (see Fig.~\ref{na24_xsec}). This strong correlation among the three systems has been well explained by a coupled channel calculations and the significantly different level densities of the compound states~\cite{Jiang2013}. At sub-barrier energies, the valence neutron(s) in $^{12}$C+$^{13}$C and $^{13}$C+$^{13}$C increase the level densities of their compound states by at least one order of magnitude in comparison to $^{12}$C+$^{12}$C and result in smooth cross sections. According to eq.5 in Ref.~\cite{Jiang2013}, the upper limit of the $^{12}$C+$^{12}$C cross section is reached using the high level density limit and can be modeled consistently with the other C+C isotope systems~\cite{Jiang2013}. The average of the $^{12}$C+$^{12}$C fusion cross section is also predicted by modulating the upper limit with the averaged ratio of the level width $<$$\Gamma$$>$ and the level spacing $<$D$>$ of $^{24}$Mg~\cite{Jiang2013}. Since the effect of $<\Gamma/D>$ is not sensitive to the energy, the shape of the averaged cross section is essentially determined by the model used for the upper limit prediction. Therefore, a model well constrained by $^{12}$C+$^{13}$C is crucial for us to set an reliable upper limit and constrain the shape of the averaged cross section for $^{12}$C+$^{12}$C.

However, the large deviations among various global models exist in $^{12}$C+$^{13}$C and affect the S-factor extrapolation as shown in Fig.~\ref{fig1}(b). Very similar to $^{12}$C+$^{12}$C, the global models, CC-M3Y+Rep, S\~ao Paulo and DC-TDHF, predict a rising trend for the $^{12}$C+$^{13}$C S-factor, while the hindrance model suggests that the S-factor reaches its maximum at $E\rm_{c.m.}$=3.45$\pm$0.37 MeV~\cite{Jiang2007,Back2014}. However, the current experimental data are insufficient to test these models. Therefore, it is crucially important to extend the fusion cross section measurement of $^{12}$C+$^{13}$C down to lower energies and provide a strict test for models which can constrain on the $^{12}$C+$^{12}$C cross section.

Obtaining a precise measurement of such small fusion cross sections is a great experimental challenge due to extremely low yields and ambient backgrounds. Up to now, only $^{12}$C+$^{12}$C has been measured down to 2 nb with errors $\le$40\% while other measurements for light systems have stopped above 10 nb, which is not low enough to differentiate the various predictions~\cite{c12pt2016,c12si2018}. Notani {\it et\ al.} has extended the measurement of $^{12}$C+$^{13}$C to 20 nb~\cite{Notani2012}. The current work further extends that effort by exploiting an underground counting lab with an ultra-low background, pushing the measurement of the ${}^{12}$C(${}^{13}$C,$p$)${}^{24}$Na cross section down to 0.22(7) nb with which the $^{12}$C+$^{13}$C fusion cross section is determined to be 0.9 nb with an uncertainty $\le$ 30\%.

\section{Experiment and results}

\begin{figure}[h]
\includegraphics[scale=0.60]{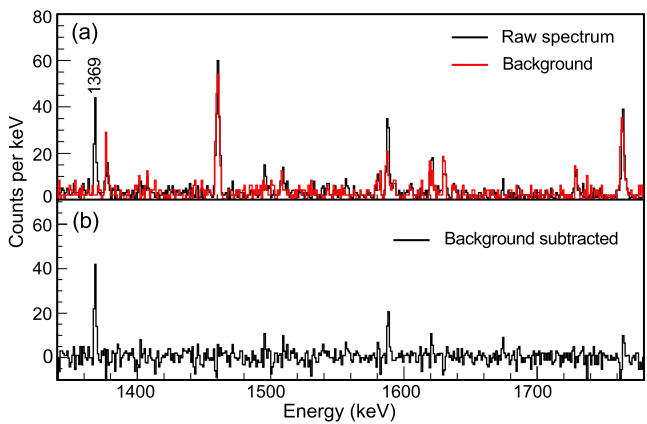}
\caption{\label{gamma}(Color online) (a) Measurements in the underground lab with 3 samples irradiated at $E\rm_{c.m.}$=2.323 MeV. The irradiations and measurements took 3.4 and 3.9 days, respectively. (b) Spectrum after subtracting background.}
\end{figure}

The fusion reaction measurement of $^{12}$C+$^{13}$C was performed by detecting the residual nucleus $^{24}$Na ($T_{1/2}$=15.0 h) from $^{12}$C($^{13}$C,p)${}^{24}$Na. A $^{13}$C beam with a typical current up to 15 p$\mu$A was delivered on a 1.5-mm-thick natural carbon target by the 3 MV Tandem of IFIN-HH~\cite{Burducea2015}. The accelerator GVM was calibrated with the $^{27}$Al(p,$\gamma$)$^{28}$Si reaction with a uncertainty less than 0.1\%. The $^{12}$C($^{13}$C,p)$^{24}$Na reaction was studied in the energy range from $E\rm_{lab}$=4.640 to 10.995 MeV with a energy step of $\Delta E_{lab}$=0.2 MeV. At lower energies the target samples were irradiated and then transported in 3 hours to the underground counting lab ($\mu$Bq) in the SLANIC salt mine, Romania for offline $\gamma$-ray measurements ~\cite{Margineanu2008}. In the measurement, a well-shielded HPGe detector was used to detect two cascading $\gamma$ rays (1369, 2754 keV) emitted by the $^{24}$Na $\beta$-decay, with only 1369 keV $\gamma$-ray being used in the final yield determination. Some of the higher energy measurements ($E\rm_{c.m.}$$>$2.8 MeV) were also performed in the Low Background Gamma-Ray Spectroscopy Laboratory (GamaSpec) in a basement of IFIN-HH. The detector efficiencies were carefully calibrated with standard sources.
The correction of the summing effect was evaluated by comparing the measurements of a ${}^{24}$Na sample at two different distances, 15 cm from the detector and a very close geometry (less than 1 mm from the detector). The results are also compared with the Monte Carlo simulations. The total times for irradiation and measurements were 20 and 33 days, respectively. Fig.~\ref{gamma} shows the $\gamma$-ray spectrum obtained at the energy of $E\rm_{c.m.}$=2.323 MeV with a ${}^{24}$Na production cross section of 0.2 nb. The 1369 keV peak, used in our yield determinations, is clearly seen in both raw and background subtracted spectra.

The fusion cross sections of $^{12}$C($^{13}$C,$p$)$^{24}$Na are obtained by differentiating the thick target yield using the procedure given in Ref.~\cite{Notani2012}. Our measurement includes a 10\% systematic uncertainty which primarily results from the uncertainties in the beam current (5\%), beam energy (2\%), detector efficiency (5\%), and stopping power (7\%) from Ref.~\cite{srim}. The measurement by Notani $\it et\ al.$ is a relative measurement. Their $^{12}$C($^{13}$C,$p$)$^{24}$Na cross sections are normalized to our data. Both results are shown in Fig.~\ref{na24_xsec}.

Two different methods, in-beam characteristic $\gamma$ ray~\cite{Dayras1976} and summing-$\gamma$-ray yield~\cite{Dasmahapatra1982-p257}, have been used to obtain the total fusion cross sections with the aid of the statistical model, making $^{12}$C+$^{13}$C a good case to assess the model$^\prime$s systematic error. The ratio between these two measurements is 1.11$\pm$0.07($\sigma$). The deviation of 0.11 from 1 and the fluctuation of 0.07 result from a combination of the statistical error and the systematic uncertainties of both the experiment and the statistical model. As a conservative value, we combine the central deviation (0.11) with the standard deviation (0.07) and estimate the systematic error of the statistical model as 13\%. This error has been included in both total cross sections in the latter part of this paper.

\begin{figure}[h]
\includegraphics[scale=0.44]{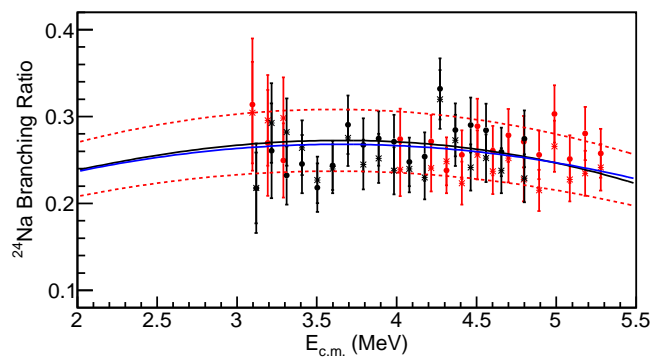}
\caption{\label{fig4:na24_br}(Color online) Branching ratio of the $^{12}$C($^{13}$C,$p$)$^{24}$Na reaction channel. The red and black symbols correspond to our measurement and the normalized Notani $et$ $al$ measurement~\cite{Notani2012}, respectively, with the branching ratios deduced from the total fusion cross sections of Ref.~\cite{Dayras1976} (circles) and Ref.~\cite{Dasmahapatra1982-p257} (stars). Solid lines show the normalized theoretical branching ratios calculated with Empire (black) and Talys (blue), with dashed lines representing the 1$\sigma$ limits from the Talys calculation.}
\end{figure}

The total fusion cross sections over the full energy range measured in this work are converted from the measured cross sections of the $^{12}$C($^{13}$C,$p$)$^{24}$Na channel based on the statistical model calculations, and hence a reliable branching ratio of the proton channel with a quantified uncertainty is needed. The experimental branching ratio is obtained by comparing the $^{12}$C($^{13}$C,$p$)$^{24}$Na cross sections to the two sets of measured total fusion cross sections~\cite{Dayras1976, Dasmahapatra1982-p257} (Fig.~\ref{fig4:na24_br}). Two different statistical model codes, Talys~\cite{talys} and Empire~\cite{empire}, are used to calculate the branching ratio (see Fig.~\ref{fig4:na24_br}). In the Talys calculation, the spin population of the ${}^{25}$Mg compound is set using the prediction by~\cite{Esbensen2011}. The potential parameters of the $p$ and $n$ channels are tuned to reproduce the partial cross sections in Ref.~\cite{Dayras1976}. In the Empire calculation, the $^{24}$Na channel branching ratio is predicted using the default parameters and normalized to the experimental data by a factor of 0.84. These two calculations are similar with a difference less than 7\%. The fluctuation of data around the best fit is 14\%(1$\sigma$), which is related to the systematic uncertainties of the measurements and statistical model. In this work, the Talys result is used for the branching ratio correction. By including the systematic difference of the two models ($\pm$3.5\%), we adopt 14\% as the systematic error for the $^{12}$C($^{13}$C,$p$)$^{24}$Na branching ratio obtained from the model. This error was overestimated as 20\% by Notani $\it et\ al.$~\cite{Notani2012}.

\section{Discussion}

\begin{figure}[h]
\includegraphics[scale=0.43]{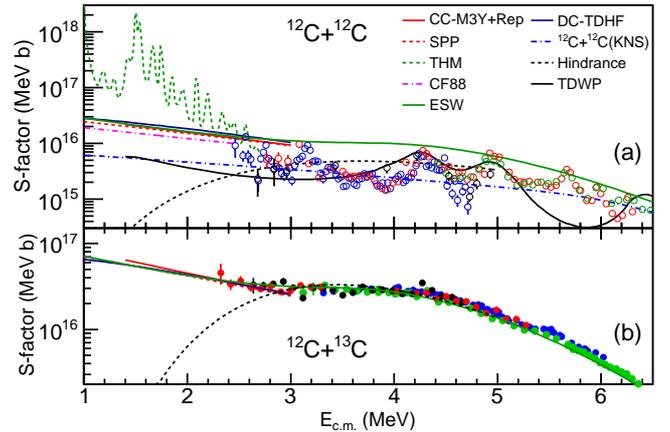}
\caption{\label{fig5:new_sfactor}(Color online) S-factors of $^{12}$C+$^{12}$C and $^{12}$C+$^{13}$C at sub-barrier energies. The data obtained by this work are shown as red filled circles. The data from Ref.~\cite{Notani2012} are normalized to the current result and are shown as black filled circles. Other symbols are identical to those in Fig.~\ref{fig1}. An uncertainty of 14\% has been added to all $^{12}$C+$^{13}$C data to account for the systematic uncertainty of the statistical model. The $^{12}$C+$^{13}$C predictions by CC-M3Y+Rep, DC-TDHF, SPP are normalized to our data below 3 MeV. The same normalization factors are also applied to the $^{12}$C+$^{12}$C accordingly. The $^{12}$C+$^{12}$C upper limit is shown as the solid green line.}
\end{figure}

The experimental S-factors together with predictions are shown in Fig.~\ref{fig5:new_sfactor}(b). A S-factor plateau, instead of a S-factor maximum at 3.45$\pm$0.37 MeV as predicted by the hindrance model, is observed in the range of 3 to 4 MeV. As energy decreases, the S-factor continues to increase as predicted by other models, SPP, CC-M3Y+Rep and DC-TDHF. The three global theoretical models shown in Fig.~\ref{fig1}(b), CC-M3Y+Rep, DC-TDHF and SPP, are higher than our result with deviations up to 35\%, 55\% and 55\%, respectively, at $E\rm_{c.m.}$$<$ 3 MeV. After normalizing these three models to our data below 3 MeV, the reduced-$\chi^2$s within this range are about 0.33. The predictions at $E\rm_{c.m.}$$=$1 MeV are consistent with each other with $\pm$12\% deviations as shown in Fig.~\ref{fig5:new_sfactor}(b). The same normalization factors of the three models are also applied to the $^{12}$C+$^{12}$C predictions (Fig.~\ref{fig5:new_sfactor}(a)).

Hindrance is a global phenomenological model predicting a rapid drop of the S-factor at lower energies for both $^{12}$C+$^{12}$C and $^{12}$C+$^{13}$C.
If this model was correct for $^{12}$C+$^{12}$C, such a strong hindrance signature would have shown also in $^{12}$C+$^{13}$C. But our results show clearly that the deviation between the experimental data and the hindrance model prediction increases at lower energies, resulting in a reduced-$\chi^2$ of 3.05 at energies below 3 MeV. At the three lowest energies, the deviation between the experimental data and the hindrance model prediction ranges from 2.3 up to 2.5$\sigma$. This observation is the first decisive experimental evidence showing that the hindrance model with the current systematics fails to be a good predictive model in the carbon isotope systems.

Although only some of the tested models can explain the observed hindrance signature in the medium heavy system~\cite{Simenel2017,Miifmmode2006,Miifmmode2007}, they all predict the exactly same shape of the $^{12}$C+$^{13}$C S-factor at deep sub-barrier energies which is confirmed by the current experiment. It suggests that the hindrance effect is less important in the carbon isotope systems. Very recently TDHF calculation also claims the absence of hindrance in $^{12}$C+$^{12}$C~\cite{PhysRevC.100.024619}. A similar absence of hindrance has also been found in a heavier system $^{7}$Li+$^{198}$Pt~\cite{c12pt2016}.

The Equivalent Square Well (ESW) is a simple model consisting of 3 parameters, reduced radius, real and imaginary potential. The prediction of this model based on the Dayras $et$ $al$. measurement provides an excellent fit to the old data sets as well as demonstrating its excellent predictive power by agreeing with our new measurements below 2.8 MeV where the minimum cross section is about 3 orders of magnitude lower. The reduced-$\chi^2$ of the fit to all the $^{12}$C+$^{13}$C data in Fig. \ref{fig5:new_sfactor}(b) is only 0.69.

After ruling out the possibility of the greatly suppressed S-factor predicted by the hindrance model and confirming the rising trend predicted by the other three global models, we introduce a new upper limit for the $^{12}$C+$^{12}$C fusion cross sections at stellar energies based on the correlation between the maxima of the $^{12}$C+$^{12}$C resonance-like structure and the $^{12}$C+$^{13}$C cross sections~\cite{Notani2012,Jiang2013}. Impressed by the excellent predictive power of ESW, we use the ESW parameters recommended for $^{12}$C+$^{13}$C to predict the $^{12}$C+$^{12}$C upper limit. This prediction provides an excellent upper bound to all the existing data shown in Fig.~\ref{fig5:new_sfactor}(a). At $E\rm_{c.m.}$$<$ 3 MeV, the three normalized global models, CC-M3Y+Rep, SPP and DC-TDHF exhibit similar S-factor shape with difference less than 20\%.

It was claimed that, within the three models tested in~\cite{Jiang2018}, Fowler (constant S* factor), CC-M3Y+Rep modulated by $<$$\Gamma/D$$>$ and the hindrance model, the hindrance model provided the best description of the average behavior of the cross section based on the $\chi^2$ test. Considering each model cannot predict the absolute value precisely, we re-fit Spillane's data using the three models in the range of 2.68 to 3.98 MeV with tunable normalization factors. The reduced-$\chi^2$s for the Fowler, the modulated CC-M3Y+Rep and the hindrance model are reduced from 502 to 18, from 1296 to 23 and from 384 to 33, respectively. This observation indicates that the resonance-like structure of $^{12}$C+$^{12}$C can not tell which is the best extrapolating model. But decisive conclusion can be made and a more reliable upper limit is established based on our $^{12}$C+$^{13}$C experiment.

We notice that the THM result is significantly higher than our new upper limit as shown in Fig.~\ref{fig5:new_sfactor}(a). There is an on-going debate~\cite{akram_pt,PhysRevC.99.064618,aurora} about how to extrapolate the THM measurements down to the stellar energies. Both accurate direct measurements at energies below $E\rm_{c.m.}$=2.7 MeV and the improvement in the THM theory are useful to resolve the tension.

Our constrains on the $^{12}$C+$^{12}$C play important roles in the astrophysical studies. The dramatic difference between the standard rate and the rate based on the hindrance model is one of the largest uncertainties in various stellar models. Using the hindrance rate, the type Ia supernovae and superburst models would ignite at higher density and temperature or even fail to explode; massive stars could produce more $^{26}$Al and $^{60}$Fe. Our result rules out these possibilities and supports the models using the standard rate. The superburst model requires strong resonances around 1.5 MeV to enhance the rate by more than 25 times and ignite the carbon burning at 0.5 GK~\cite{Cooper2009}. With our upper limit, the maximum enhancement is about a factor 2 instead. It suggests more physics needs to be included in the current superburst model to solve the ignition problem.

\section{Conclusion}

As a conclusion, the strong correlation between the $^{12}$C+$^{12}$C and $^{12}$C+$^{13}$C fusion cross sections offers a great opportunity to constrain the upper limit for the $^{12}$C+$^{12}$C fusion cross sections at lower energies. However, such an upper limit towards the stellar energies suffers from the large deviations among various models. Our new measurement of the $^{12}$C+$^{13}$C fusion cross section down to $E\rm_{c.m.}$=2.323 MeV disagrees with the prediction of the hindrance model and is the first decisive evidence in the carbon isotope systems that rules out the existence of the S-factor maximum predicted for $^{12}$C+$^{13}$C by this phenomenological model. It also confirms the exponentially rising trend of the S-factor towards lower energies predicted by CC-M3Y+Rep, DC-TDHF, KNS, SPP and ESW. After calibrating these model predictions using our data, a more reliable upper limit is established for the $^{12}$C+$^{12}$C fusion cross sections at stellar energies.

\section{Acknowledgments}
X.D.T. thanks H. Esbensen, C.Y. Wong, A. Mukhamedzhanov, D.Y. Pang, X. Fang, W.P. Tan, S.F. Zhu, D.L. Fang and X.D. Xu for the helpful discussions, A. Tumino for providing the THM data and A. Tang for proofreading. This project is supported in part by the National Key Research and Development program (MOST 2016YFA0400501) from the Ministry of Science and Technology of China and the grants PN 16420102 and PN-III-P4-ID-PCE-2016-0743 NUCASTRO2 from the Romanian Ministry for Research and Innovation.
X.D.T. and N.T.Z. acknowledge supports from the National Natural Science Foundation of China under Grants No. 11021504, No. 11321064, No. 11475228, No. 11490564, and No. 11405226. X.D.T. also acknowledges supports from the key research program (XDPB09-2) and 100 talents Program of the Chinese Academy of Sciences.
B.B. and S.U. acknowledge support by the U.S. DOE under Contract No. DE-AC07-05ID14517 and Grant No. DE-SC0013847.

\section*{References}
\bibliography{13C12C_ref_v7}







\end{document}